\documentclass[twocolumn]{aastex7}
\usepackage{epsfig,graphicx,natbib}
\usepackage{hyperref,url}

\usepackage{amsmath}
\usepackage{natbib}
\usepackage{xspace}
\usepackage[varg]{txfonts}
\usepackage{graphicx}
\usepackage{nicefrac}
\usepackage{tablefootnote}
\usepackage{multirow}
\usepackage{soul}
\usepackage{float}
\usepackage[T1]{fontenc}
\usepackage{lmodern}
\usepackage{multirow}
\usepackage{ulem}
\usepackage{soul}

\graphicspath{{./figures/}{./}}

\newcommand{\Msun}{\ensuremath{\,M_\odot}\xspace}

\begin{document}

\title{HST Deep Upper Limits Rule Out a Surviving Massive Binary Companion to the Type Ic Supernova 2012fh}

\author[0000-0002-7502-0597]{Benjamin F. Williams}
\affiliation{Department of Astronomy, University of Washington, Box 351580, Seattle, WA 98195-1580, USA}
\email{benw1@uw.edu}

\author[0000-0002-7464-498X]{Emmanouil Zapartas}
\affiliation{Institute of Astrophysics, FORTH, N. Plastira 100,  Heraklion, 70013, Greece}
\email{ezapartas@ia.forth.gr}

\author[0000-0003-2238-1572]{Ori D. Fox}
\affiliation{Space Telescope Science Institute, 3700 San Martin Drive, Baltimore, MD 21218, USA}
\email{ofox@stsci.edu}

\author[0000-0002-4924-444X]{K.\ Azalee Bostroem}
\affiliation{Steward Observatory, University of Arizona, 933 North Cherry Avenue, Tucson, AZ 85721-0065, USA}
\altaffiliation{LSST-DA Catalyst Fellow}
\email{bostroem@arizona.edu}

\author[0000-0003-0607-8194]{Jianing Su}
\affiliation{David A.~Dunlap Department of Astronomy \& Astrophysics, University of Toronto, 50 St.~George Street, Toronto, ON M5S 3H4, Canada}
\email{jennyjn.su@mail.utoronto.ca}

\author[0000-0001-5530-2872]{Brad Koplitz}
\affiliation{School of Earth \& Space Exploration, Arizona State University, 781 Terrace Mall, Tempe, AZ 85287, USA}
\email{bkoplitz@asu.edu}

\author[0000-0001-9038-9950]{Schuyler D.~Van Dyk}
\affiliation{Caltech/IPAC, Mailcode 100-22, Pasadena, CA 91125, USA}
\email{vandyk@ipac.caltech.edu}

\author[0000-0001-7081-0082]{Maria R. Drout}
\affiliation{David A.~Dunlap Department of Astronomy \& Astrophysics, University of Toronto, 50 St.~George Street, Toronto, ON M5S 3H4, Canada}
\email{maria.drout@utoronto.ca}

\author[0000-0002-6548-5489]{Dimitris Souropanis}
\affiliation{Institute of Astrophysics, FORTH, N. Plastira 100,  Heraklion, 70013, Greece}
\email{dsouropanis@ia.forth.gr}

\author[0000-0002-0763-3885]{Dan Milisavljevic}
\affiliation{Department of Physics and Astronomy, Purdue University, 525 Northwestern Avenue, West Lafayette, IN 47907, USA}
\affiliation{Integrative Data Science Initiative, Purdue University, West Lafayette, IN 47907, USA}
\email{dmilisav@purdue.edu}

\author[0000-0003-4501-8100]{Stuart D. Ryder}
\affiliation{School of Mathematical and Physical Sciences, Macquarie University, NSW 2109, Australia}
\affiliation{Astrophysics and Space Technologies Research Centre, Macquarie University, Sydney, NSW 2109, Australia}
\email{stuart.ryder@mq.edu.au}

\author[0000-0001-9336-2825]{Selma~E.~de~Mink}
\affiliation{Max Planck Institute for Astrophysics, Karl-Schwarzschild-Straße 1, 85748 Garching b. M{\"u}nchen, Germany}
\email{sedemink@gmail.com}

\author[0000-0001-5510-2424]{Nathan~Smith}
\affiliation{Steward Observatory, University of Arizona, 933 North Cherry Avenue, Tucson, AZ 85721-0065, USA}
\email{nathans@as.arizona.edu}

\author[0000-0001-8416-4093]{Andrew Dolphin}
\affiliation{Steward Observatory, University of Arizona, 933 North Cherry Avenue, Tucson, AZ 85721-0065, USA}
\affiliation{Raytheon Technologies, 1151 East Hermans Road, Tucson, AZ 85706, USA}
\email{adolphin@rtx.com}

\author[0000-0003-3460-0103]{Alexei V. Filippenko}
\affiliation{Department of Astronomy, University of California, Berkeley, CA 94720-3411, USA}
\email{afilippenko@berkeley.edu}

\author[0000-0001-5261-3923]{Jeff J.\,Andrews}
\affiliation{Department of Physics, University of Florida, 2001 Museum Rd, Gainesville, FL 32611, USA}
\email{jeffrey.andrews@ufl.edu}

\author[0000-0002-6842-3021]{Max M.\,Briel}
\affiliation{Département d'Astronomie, Université de Genève, Chemin Pegasi 51, CH-1290 Versoix, Switzerland}
\affiliation{Gravitational Wave Science Center (GWSC), Université de Genève, CH1211 Geneva, Switzerland}
\email{max.briel@gmail.com}

\author[0000-0001-6692-6410]{Seth Gossage}
\affiliation{Center for Interdisciplinary Exploration and Research in Astrophysics (CIERA), Northwestern University, 1800 Sherman Ave, Evanston, IL 60201, USA}
\email{seth.gossage@northwestern.edu}

\author[0000-0001-9331-0400]{Matthias U.\,Kruckow}
\affiliation{Département d'Astronomie, Université de Genève, Chemin Pegasi 51, CH-1290 Versoix, Switzerland}
\affiliation{Gravitational Wave Science Center (GWSC), Université de Genève, CH1211 Geneva, Switzerland}
\email{Matthias.Kruckow@unige.ch}

\author{Camille Liotine}
\affiliation{Department of Physics and Astronomy, Northwestern University, 2145 Sheridan Road, Evanston, IL 60208, USA}
\affiliation{Center for Interdisciplinary Exploration and Research in Astrophysics (CIERA), Northwestern University, 1800 Sherman Ave, Evanston, IL 60201, USA}
\email{cliotine@u.northwestern.edu}

\author[0000-0003-1749-6295]{Philipp M.\,Srivastava}
\affiliation{Center for Interdisciplinary Exploration and Research in Astrophysics (CIERA), Northwestern University, 1800 Sherman Ave, Evanston, IL 60201, USA}
\affiliation{Electrical and Computer Engineering, Northwestern University, 2145 Sheridan Road, Evanston, IL 60208, USA}
\affiliation{NSF--Simons AI Institute for the Sky (SkAI), 172 E.\,Chestnut St., Chicago, IL 60611, USA}
\email{elizabethteng@u.northwestern.edu}
\email{Philipp.msrivastava@northwestern.edu}

\author{Elizabeth Teng}
\affiliation{Department of Physics and Astronomy, Northwestern University, 2145 Sheridan Road, Evanston, IL 60208, USA}
\affiliation{Center for Interdisciplinary Exploration and Research in Astrophysics (CIERA), Northwestern University, 1800 Sherman Ave, Evanston, IL 60201, USA}
\affiliation{NSF--Simons AI Institute for the Sky (SkAI), 172 E.\,Chestnut St., Chicago, IL 60611, USA}
\email{elizabethteng@u.northwestern.edu}

\begin{abstract}

Current explanations of the mass-loss mechanism for stripped-envelope supernovae remain divided between single and binary progenitor systems. Here we obtain deep ultraviolet (UV) imaging with the {\it Hubble Space Telescope (HST)} of the Type Ic SN 2012fh to search for the presence of a surviving companion star to the progenitor. We synthesize these observations with archival {\it HST} imaging, ground-based spectroscopy, and previous analyses from the literature to provide three independent constraints on the progenitor system. We fit the color-magnitude diagram of the surrounding population to constrain the most likely age of the system to be $<20$\,Myr.  Analysis of spectra of SN~2012fh provide an estimate of the He core mass of the progenitor star, $>5.6$ M$_{\odot}$.  We analyze deep {\it HST} images at the precise location after the SN faded to constrain the luminosity of any remaining main-sequence binary companion to be $\log(L/L_{\odot}) \lesssim 3.35$. Combining  observational constraints with current binary population synthesis models excludes the presence of a faint stellar companion to SN~2012fh at the $\lesssim10\%$ level. The  progenitor was therefore either effectively isolated at the time of explosion or orbited by a black-hole companion.  The latter scenario dominates if we only consider models that produce successful supernovae.
\end{abstract}

\section{Introduction}\label{sec:intro}

Most massive stars end their lives when their core runs out of fuel for nuclear fusion and collapses, causing a core-collapse supernova (SN) explosion \citep[see][for a review]{jerkstrand2025}.  Some of these explosions are observed to be deficient in hydrogen, as their spectra show weak or no hydrogen lines.  Such supernovae (SNe) are classified as Type Ib, Ic, and IIb, and are thought to arise from progenitors that have lost most or all of their hydrogen-rich envelope prior to explosion; these events are therefore called stripped-envelope supernovae (SESNe).  One likely mechanism by which a massive star could lose the hydrogen surrounding its core is through interactions with a binary companion \citep[e.g.,][]{Podsiadlowski1992,eldridge2008,yoon2017,zapartas2017b}, which may have already collapsed to form a neutron star (NS) or black hole (BH).  Another mechanism is the standard single-star scenario, where massive stars produce stripped-envelope Wolf-Rayet (WR) progenitors via continuous stellar winds \citep{heger03} or luminous blue variable eruptions \citep{so06}.  However, modern estimates of wind mass-loss rates have been significantly reduced \citep{smith14}, making it difficult to remove the H envelope across most stellar masses and impossible to produce enough SESNe via winds alone \citep{smith11frac}.  SESNe require a substantial contribution from binary stripping, but may result from a mix of both binary stripped progenitors and single wind-stripped progenitors.  

Significant effort has been made to constrain the physical nature of the progenitors of SESNe.  There have been direct detections of the progenitors of SNe IIb (e.g., SN 1993J, \citealp{aldering1994}; SN 2011dh, \citealp{strotjohann2015}), two SNe Ib (iPTF13bvn, \citealp{cao2013}; SN 2019yvr, \citealp{kilpatrick2021}), and one SN Ic candidate (SN 2017ein, \citealp{vandyk2018,Zhao2025}).  Their spectra have put constraints on explosion energies, ejecta masses, and even progenitor masses \citep[e.g.,][]{mazzali2007,teffs2021,teffs2022}.  Deep searches for the remaining binary companions have resulted in the detection of several candidates for Type IIb and Ib events \citep[e.g.,][]{vandyk2002,Maund2004,maund2016,ryder2006,ryder2018}, but only one candidate has been detected for an SN~Ic \citep[SN 2013ge;][]{fox2022}. Furthermore, deep upper limits on the presence of any surviving companion to the SNe Ic 2002ap \citep{zapartas2017b} and 2008ax \citep{folatelli15} rule out the presence of a massive $>8$--10 \Msun\ main-sequence (MS) companion star (although lower mass companions were not ruled out). Yet analysis of the brightest stars close to the sites of several events suggest that the progenitors of Type~Ic SESNe may be younger on average than those of other types of SESNe \citep{maund2018,sun2023}, making SN~Ic progenitors likely to be particularly massive.

SN~2012fh was a Type Ic SN in NGC 3344 \citep{nakano2012,takaki2012,margutti2012} at a distance of 9.8 Mpc \citep{jacobs2009,anand2021} which is assumed throughout this work, making it one of the most nearby SNe~Ic ever observed.  The SN was not observed at its peak brightness owing to the proximity of the Sun \citep{johnson2017}.  It was first detected on 2012 Oct. 18 (UTC dates are used throughout this paper) by \citet{nakano2012} with an apparent magnitude of 15.1 (unfiltered), and classified as an SN~Ic at about 130 days past maximum luminosity.  Fortunately, multi-epoch ground-based imaging covers the location prior to the explosion \citep{gerke2015,johnson2017}.  
Photometric limits were placed at the location of SN~2012fh from Large Binocular Telescope imaging taken prior to the event.  These measurements are detailed by \citet{johnson2017}. 
Furthermore, we have followed up the event with deep {\it Hubble Space Telescope (HST)} imaging, which we  include along with archival {\it HST} images from previous monitoring, to constrain any possible massive stellar companion that would still be at the SN location after the event faded.  

In this paper, we present results from deep ultraviolet (UV) {\it HST} observations of SN 2012fh to search for the presence of a surviving companion to its progenitor. Deep UV imaging is most sensitive to the presence of any surviving MS massive companion star. We fully analyze the photometry of the SN location before and after the explosion, as well as the photometry of the stars in the surrounding region and spectroscopy of the event itself, to place as many constraints on the physical parameters of the progenitor system as possible.  In Section 2, we describe all of the data we apply to our investigation, as well as our analysis techniques.  Section 3 details the results from our analysis, including the mass and age constraints on the progenitor binary system.  We discuss in Section 4 the implications of these results on SESN theory, and Section 5 provides a summary of our conclusions.

\section{Data and Analysis}

We analyzed both images and spectra in our study of SN 2012fh.  The imaging was all obtained by {\it HST}.
Table \ref{tab:tab1} summarizes the observations obtained of SN 2012fh with the {\it HST} WFC3/UVIS channel as part of program GO-16165 (PI O. Fox), as well as all archival data that cover that location (GO-13364, PI D. Calzetti) from 2014 March 2 and GO-13773 (PI R. Chandar) from 2015 March 1, as well as from GO-14762 (PI J. Maund) from 2016 December 25. While there is also imaging of NGC 3344 from February of 2012 (GO-12546; PI B. Tully), before the SN, it unfortunately did not include the SN location.  Thus, the pre-supernova constraints come from the literature.  In the following subsections, we briefly discuss these limits, followed by our new post-supernova photometry and spectroscopy.

\subsection{Pre-supernova Photometry Limits}
\cite{johnson2017} found upper limits on the brightness of the progenitor $M_U>-3.8$, $M_B>-3.1$, $M_V >-3.8$, and $M_R > -4.0$\,mag.  They assumed a distance of 6.9 Mpc for these limits.  The equivalent limits for our assumed distance are then $M_U>-4.6$, $M_B>-3.9$, $M_V >-4.6$, and $M_R > -4.8$\,mag.  

\subsection{Imaging Analysis}

The individual UVIS {\tt flc} frames in all bands were obtained from the Barbara A. Mikulski Archive for Space Telescopes (MAST)\footnote{ \url{https://doi.org/10.17909/zarq-p121}}. We note that there are two sets of observations of SN 2012fh for GO-16165. During the first set (2021 Jan. 13), several exposures were flagged as poor/unusable owing to telescope tracking issues. Following a HOPR request, all observations were obtained again on 2022 Jan. 02. Although spaced nearly one year apart, the relative separation compared to the age of the SN ($\sim 10$\,yr) is small. Furthermore, the SN was not detected at either epoch. For these reasons, we decided to combine the two sets of exposures, keeping all exposures from the first epoch that were not affected by the tracking issues. This has the benefit of resulting in deeper imaging than originally proposed. 
We aligned all images using {\tt jhat}\footnote{https://github.com/arminrest/jhat}. The frames in each band then had cosmic-ray hits masked and were combined into mosaics by running them through AstroDrizzle in PyRAF. 

We conducted two different types of analysis on the images.  First, we ran photometry forced on only the pixels at the location of the SN to measure the brightness of the fading event and upper limits of any remaining companion.  Then, we ran crowded-field point-spread-function (PSF) fitting photometry on the field surrounding the location of the SN, which probes the population of young stars likely to be associated with the progenitor.  We describe each of these processes in the following subsections. 

\subsection{Supernova Location Photometry}

\begin{figure*}
  \centering
    \includegraphics[height=70mm]{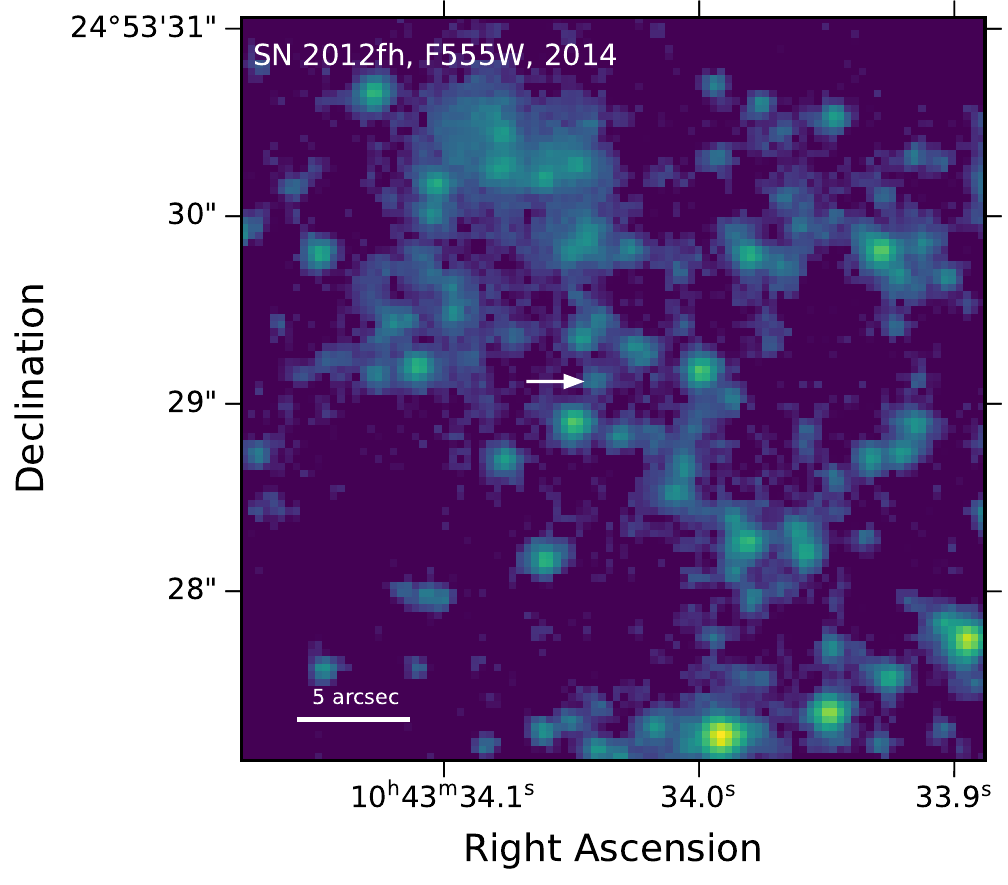}  
  \includegraphics[height=70mm]{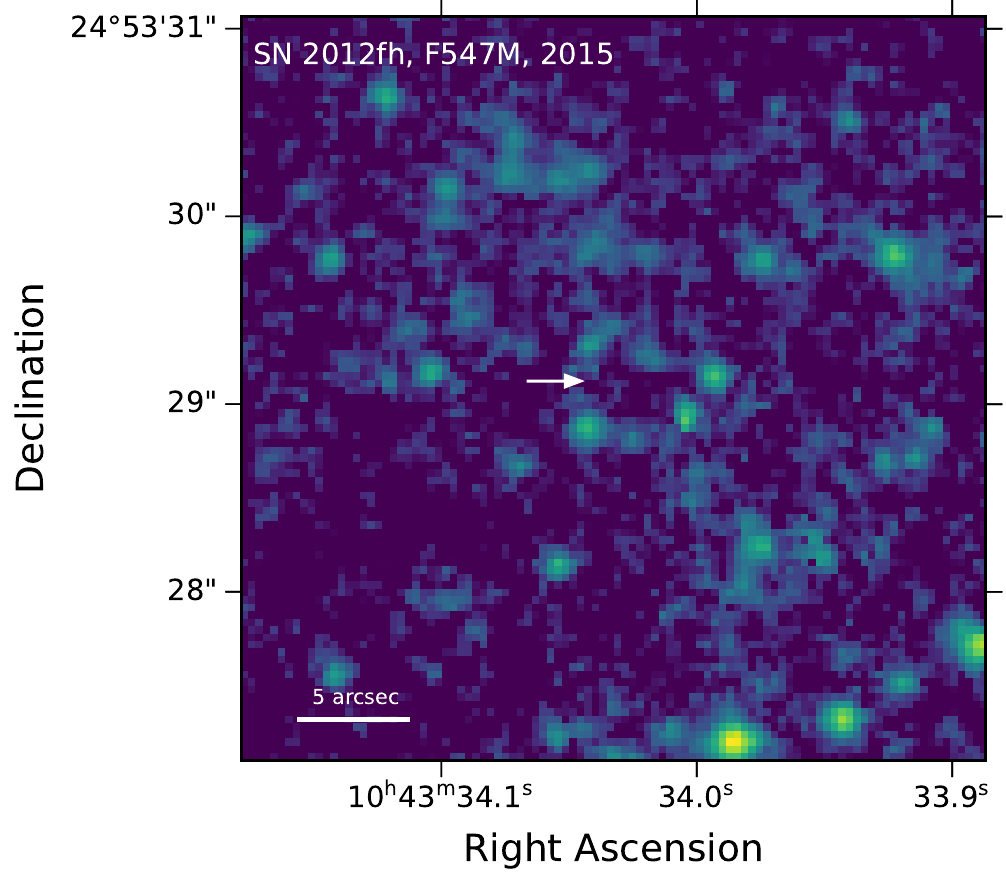}  
  \includegraphics[height=70mm]{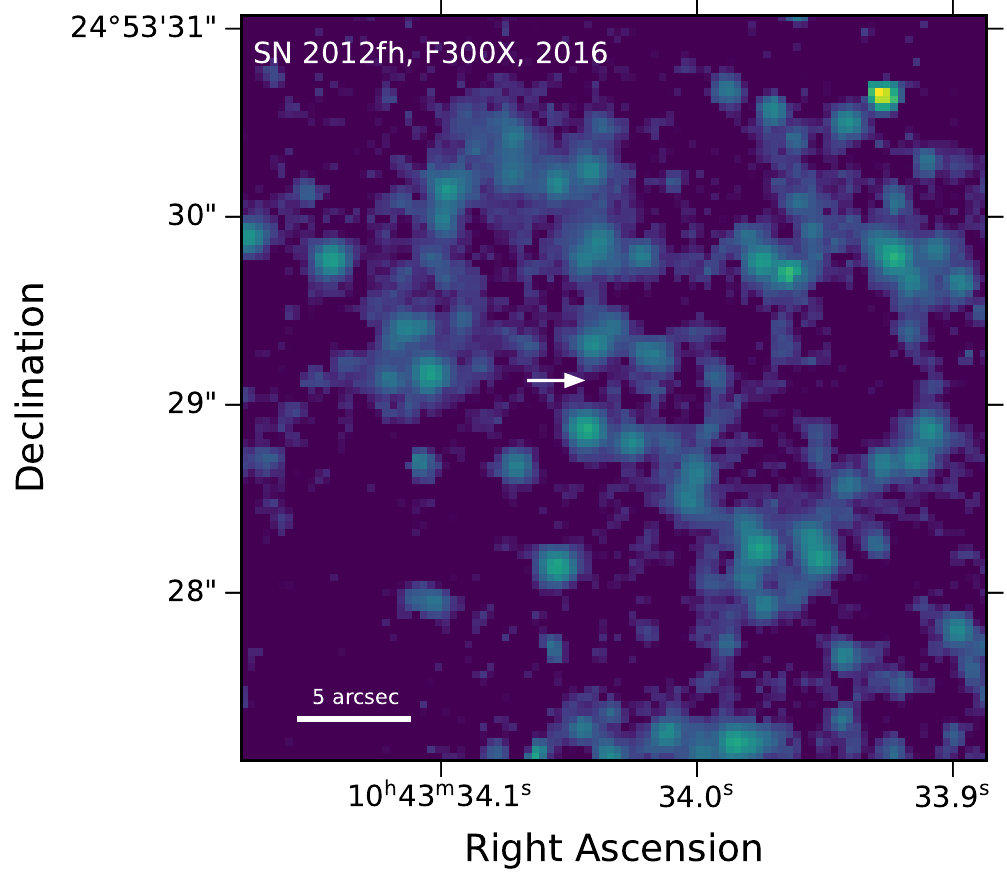}  
  \includegraphics[height=70mm]{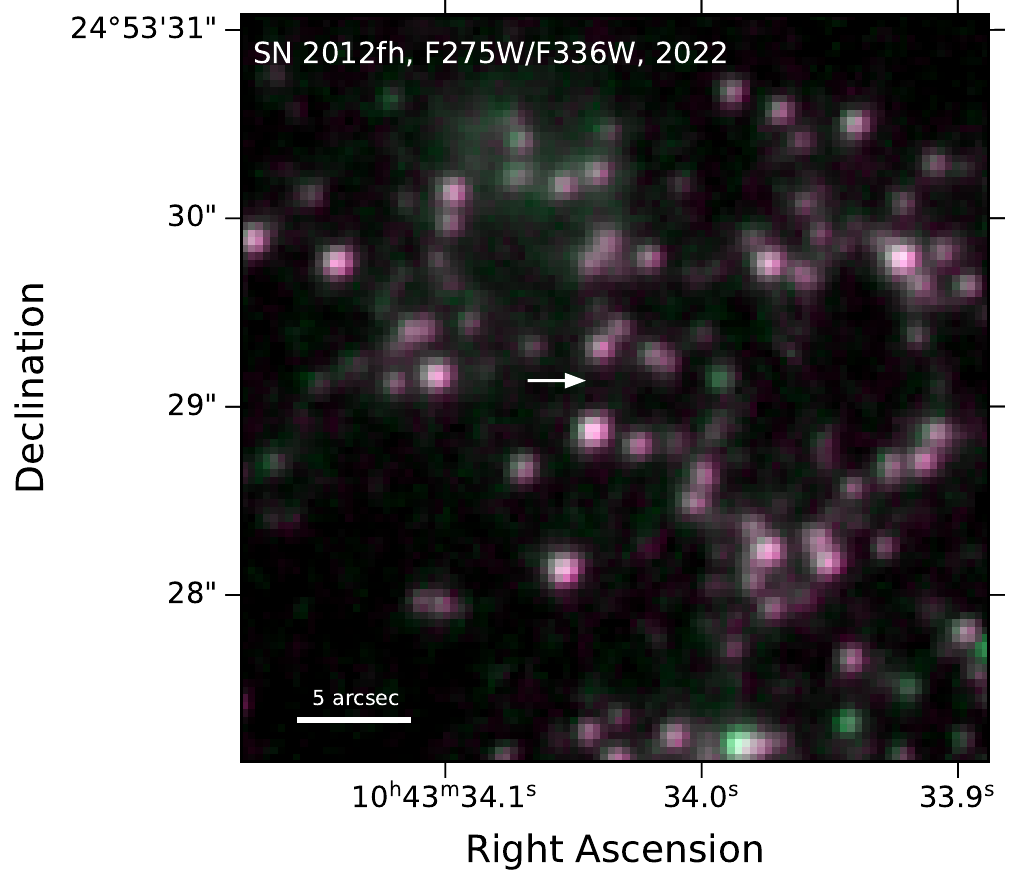}  
  \caption{Post-explosion images of SN 2012fh obtained from the MAST {\it HST} archive. {\it Upper left:} The SN location is confirmed in F555W images from 2014 (GO-13364), where the fading event is still visible.  {\it Upper right:} The SN is no longer seen in F547M images from 2015 (GO-13773).  {\it Lower left:} The SN is no longer seen in F300X images from 2016 (GO-14762). {\it Lower right:} False-color image (F275W = Blue, F336W = Red) of our 2022 UVIS imaging of the same location.  In all images, the position of the SN is noted by the arrow.  In all but the 2014 imaging, no source is visible, suggesting any remaining binary companion is not a very massive star. We derive upper limits listed in Table \ref{tab:tab1}. Five arcsec is 240~pc at our assumed distance to SN~2012fh.
  }
\label{fig:post_color}
\end{figure*}

Our SN photometry from the imaging is dependent on accurately determining the SN position.  We searched for a surviving companion in the stacked {\tt drz} files by aligning the new data to archival data where the SN is clearly detected and the position is well known. Again, we used {\tt jhat} to complete this alignment. Figure \ref{fig:post_color} shows a color composite of our data compared to 2014 imaging of the field with the F555W filter (GO-13364; PI D. Calzetti). 

\begin{figure}
  \centering
  \includegraphics[width=90mm]{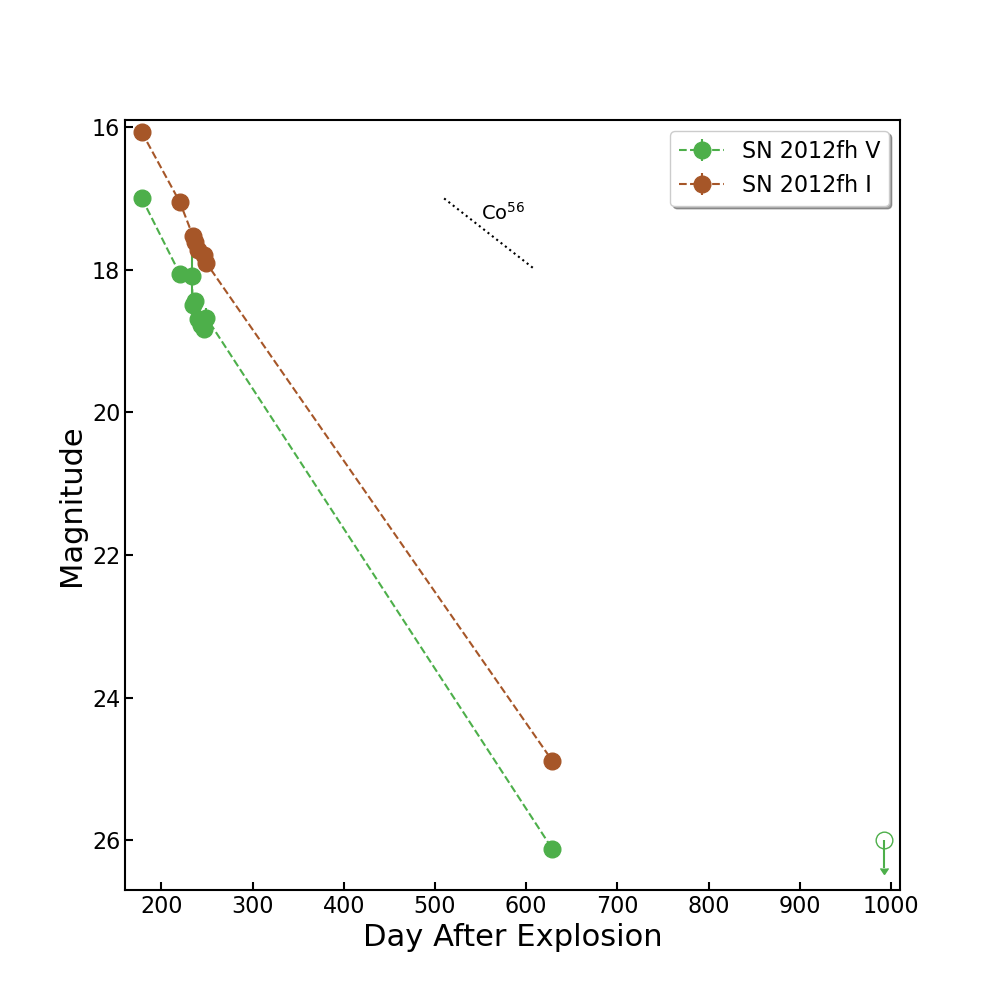}  
  \caption{Post-explosion photometry of SN 2012fh in $V$ and $I$ from \citet{2022Zheng}, together with measurements from archival {\it HST} F555W ($\sim V$) and F814W ($\sim I$) late-time images from 2014 (GO-13364; PI D. Calzetti); see Table \ref{tab:tab1}.  We adopt an explosion date of 2012 June 12.13, to match that adopted for the spectroscopy.  The fact that the late-time measurements are steeper than radioactively powered exponential decline is common for SNe Ib/c and likely is due to incomplete trapping of gamma-rays caused by the relatively low ejecta mass.  Most importantly, the source was no longer detectable in 2016 in {\it HST} F547M, provides us with confidence that the source, indicated in Figure \ref{fig:post_color}, is SN 2012fh.
  }
\label{fig:sn12fh_lc}
\end{figure}

We are highly confident that we have isolated the SN position, since a clearly detectable source is seen and measured in the 2014 data; see Figure \ref{fig:post_color} and Table \ref{tab:tab1}. This source is not detected in F547M image data from exactly one year later, in 2015, which would be in agreement with a fading SN. In fact, when we combine the 2014 F555W ($\sim V$) and F814W ($\sim I$) detections together with the SN $V$ and $I$ light curves from \citet{2022Zheng}, the late-time points are consistent with the long-term exponential decline of the SN light (Figure \ref{fig:sn12fh_lc}). This further bolsters our identification of the old SN. We note that the late-time decline rate observed is faster than expected from the radioactive decay of $^{56}$Co (as shown by the dotted line in Figure~\ref{fig:sn12fh_lc}). However, this is not uncommon for Type Ib/c SNe, and is typically attributed to incomplete trapping of gamma-rays (see, e.g., \citealt{Clocchiatti1997}).  

We used \texttt{space\_phot}\footnote{https://github.com/jpierel14/space\_phot} to perform forced PSF photometry on the DRZ mosaics for F275W, F336W, F438W, F555W, and F814W from 2014 observations (GO-13364). To account for any poor-quality pixels (from cosmic rays or faulty pixels on the detector) in our images, we used \texttt{astrodrizzle} to generate a weight (\texttt{wht}) image such that poor pixels are assigned with lower weight \citep{drizzlepac_handbook}. Then, with \texttt{create\_neg\_rms\_images} \citep{2022zndo...7458442R, 2023ApJS..265...40R}, we turned the combined weight map into an error array ultimately used by our forced PSF-fitting routine. 

Photometry of the SN was performed on the drizzled images to obtain the deepest limits. Given the complex environment, PSF photometry is most reliable. To perform PSF photometry in a mosaiced {\it HST} image, we generate a mosaiced PSF at the precise SN coordinates using \texttt{space\_phot}. Forced photometry is then performed at the SN location. During the fitting itself, we set a background noise based on a source-free region around the SN site on DS9. A detection is determined when the flux exceeds three times the uncertainty. Otherwise, we report the $5\sigma$ upper limit. We list the resulting photometry in \autoref{tab:tab1}. 


\subsection{Crowded-Field Photometry}

In addition to the \texttt{space\_phot} analysis on the specific pixels at the location of the SN, we ran crowded-field photometry on both the F275W and F336W images from GO-16165 (more than 8\,yr after the event), when any UV light from the SN had long faded, to measure the resolved stellar photometry of the surrounding stellar field.  This photometry was for the analysis of the local young stellar populations. We measured resolved stellar photometry from the {\it HST} images using the software package DOLPHOT \citep{dolphin2016} with the same photometry pipeline, including parameters and quality metrics, used and described by several other crowded-field photometry programs \citep[e.g.,][]{dalcanton2009,petia2017,williams2021,tran2023}.

In brief, the images were put through the {\tt astrodrizzle} routine to identify and flag cosmic rays. Flagged pixels are then masked, and the individual exposure images are each put into a single run of DOLPHOT, which calculates the combined noise level, as well as the observed background to find all pixels in excess of the noise.  These locations are all then fitted with the known {\it HST} PSF at the location on the detector for the band observed.  The combined measured photometry is output to a catalog, which is then post-processed to flag measurements in each band that do not pass our specific quality metrics, based on signal-to-noise ratio, sharpness, and crowding.  We note that while DOLPHOT is not designed for measuring upper limits at specific locations in the data, the depth of the DOLPHOT photometry was consistent with the \texttt{space\_phot} upper limits.

Once the initial photometry catalogs were completed, we generated artificial-star tests (ASTs) to quantify the completeness and precision of the photometry using standard techniques \citep[e.g.,][]{dalcanton2009,williams2014}. In brief, we placed a star in our region of interest into the images, and reran our photometry routine to determine if the star was recovered and, if so, the difference between the input and output magnitudes.  50,000 of these tests were run, covering the colors and magnitude ranges of the observed stars and beyond.  We include artificial stars beyond the range of detections to gauge our frequency of detecting upscattered stars, as well as to quantify our sensitivity to colors and magnitudes that we did not detect.  These statistics are then applied when fitting models to determine the stellar population age.


\begin{deluxetable*}{ l c c c c c c}
\caption{{\it HST}/WFC3 imaging of SN 2012fh \label{tab:tab1}}
\tablehead{
\colhead{Filter} & \colhead{UTC Date} & \colhead{MJD} & \colhead{Epoch} & \colhead{Exposure\tablenotemark{a}} & \colhead{SN Magnitude} & \colhead{Program}\\
\colhead{}   & \colhead{}     & \colhead{} & \colhead{(days post discovery)} &  \colhead{(s)}      & \colhead{Vega (err)} & \colhead{ID}}
\startdata
\hline
\multirow{3}{*}{F275W} & 20140302 & 56719 & 501 & 2361 & {$>$26.4} & GO-13364 \\
    &  20210113 & 59227 & 3009 & 2623 & \multirow{2}{*}{$>$26.5} & GO-16165 \\
    & 20220102 & 59581 & 3363 & 5246 &  & GO-16165 \\
\hline
F300X &  20161225  & 57748 & 1530 &  1200 & $>$26.5 & GO-14762 \\
\hline
\multirow{3}{*}{F336W} & 20140302  & 56719 & 501 &  1134 & $>$26.5 & GO-13364 \\ & 20210113  &  59227 &  3009 & 1773 & \multirow{2}{*}{$>$26.8}  & GO-16165 \\
    & 20220102 & 59581 & 3363 & 2364 & & GO-16165 \\
\hline
F438W &  20140302  & 56719 & 501 &  956 & $>$26.6 & GO-13364 \\
\hline
F475X &  20161225  & 57748 & 1530 &  350 & $>$27.5 &  GO-14762 \\
\hline
F547M &  20150301  & 57083 & 865 &  549 & $>$26.0 &  GO-13773 \\
\hline
F555W &  20140302  & 56719 & 501 &  1134 & $26.4 \pm 0.1$ & GO-13364 \\
\hline
F657N &  20150301  & 57083 & 865 &  1545 & $>$24.4 & GO-13773 \\
\hline
F814W &  20140302  &  56719 & 501 &  980 & $25.7 \pm 0.1$ & GO-13364 \\
\hline
\enddata
\tablenotetext{a}{Integration times adjusted for discarded exposures.}
\end{deluxetable*}

\subsection{Local Stellar Population}\label{sec:fitting}

To quantify the age distribution of the young stars near SN 2012fh, we constrained the recent star-formation history (SFH) of the stellar populations within a projected distance of 50\,pc from the position of SN 2012fh following the techniques used in several other similar studies \citep[e.g.,][]{badenes2009,murphy2011,jennings2014,williams2018,maund2018,koplitz2021,bostroem2023}.  Briefly, we fit the Padova single-star stellar evolution models \citep[PARSEC;][]{bressan2012} to the color-magnitude diagram (CMD) of the stars using the software package MATCH \citep{dolphin2002,dolphin2012,dolphin2013}.  It applies the AST results from our data to the models and finds the combination of model ages, within the prior-set metallicity range, that best matches the observed distribution of stars.  It also returns estimates of random uncertainties on the best-fit SFH using a hybrid Monte Carlo technique \citep{dolphin2013}.  The depth of photometry limited our sentitivity to ages younger than 50 Myr, and there is clearly a population present younger than this limit.  We limited the allowed model metallicity range for the ages of interest ($< 50$ Myr) to [Fe/H] $>-0.3$ \citep{calzetti2015,sabbi2018}. 

We also performed a grid search to find the foreground extinction, and differential extinction ($dA_V$ in the MATCH package, approximated by a top-hat distribution) to find the extinction properties that best fit  the star sample taken from within 50\,pc of location of SN 2012fh.  Finally, we use the package to perform a Monte Carlo search through the free parameters to find acceptable combinations to allow reliable estimates of the uncertainties in the age distribution \citep{dolphin2012}. 

We note that because we fit single-star models to the data, interacting binaries that have been rejuvenated though mass exchange could appear as a younger population of stars.  However, the fitting requires consistency with the rest of the observed color-magnitude distribution assuming a standard \citep{kroupa2001} initial mass function (IMF), and such rejuvenated stars will not have corresponding stars following the IMF, limiting their impact on the SFH.
The inclusion of interacting binaries in the age analysis of the surrounding region (independent of whether the SN 2012fh scenario itself involves a binary) would have allowed for older inferred ages, as binary products tend to make a region look younger \citep{Vanbeveren+1998, Wofford+2016, Xiao+2019}. Thus, this feature of the fitting would not preclude the very young ages of a few Myr required to accommodate massive WR stars.

\subsection{Spectroscopic Data}

Our spectroscopic data were from the Lick and MDM Observatories.  We obtained three nebular spectra of SN~2012fh.
Two spectra were taken with the Kast spectrograph on the Shane 3\,m telescope at Lick Observatory on 06.56 and 14.56 Nov. 2012, and were previously published by \citet{2019Shivvers};
we downloaded them from the Supernova Database (SNDB) at UC Berkeley.
A third spectrum was obtained with the Boller \& Chivens CCD Spectrograph (CCDS) on the 2.4\,m Hiltner Telescope at the MDM Observatory on 25.51 Oct. 2012. A north-south $1.5^{\prime\prime} \times 5^{\prime}$ slit and 150 lines mm$^{-1}$ grating blazed at 4700~\AA\  was used to obtain $2 \times 600$\,s exposures. Spectra were of $\sim 10$\,\AA\ resolution and span wavelengths 3450--7100\,\AA. The MDM data were processed using standard procedures including dark subtraction, flatfielding, and flux calibration in PyRAF/IRAF\footnote{IRAF is distributed by the National Optical Astronomy Observatories, which are operated by the Association of Universities for Research in Astronomy, Inc., under cooperative agreement with the U.S. National Science Foundation.} with standard stars from \citet{Strom77}.
We can also measure the properties of the SN itself from the spectra taken of the event.  
The spectra are scaled to the $BVRI$ photometry of \citet{2022Zheng} using the \texttt{lightcurve\_fitting} package \citep{2024Hosseinzadeh}.
To derive a scale factor, we integrate the spectrum convolved with the filter transmission and use the average ratio of the integrated spectrum and the actual photometry over all of the filters.

We correct the spectra for $A_{V}=0.09$\,mag based on the all-sky extinction maps of \citet{schlafly2011} and a redshift of $0.001935$ \citep{2008Epinat}. As an explosion epoch, we adopt 12.13 June 2012, 130 days before the classification spectrum was obtained \citep{nakano2012}.

\section{Results}

In this section, we report the results of the measurements of all the SN~2012fh data.  We discuss the limits on the pre-explosion properties of the progenitor, the age of the surrounding population, the mass of the core that produced the supernova, and constraints on the properties of any remaining binary companion.

\subsection{Constraints on the Progenitor from Pre-Expolosion Photometry}\label{sec:pre_explosion}

\begin{figure}
    \centering
    \includegraphics[width=\linewidth]{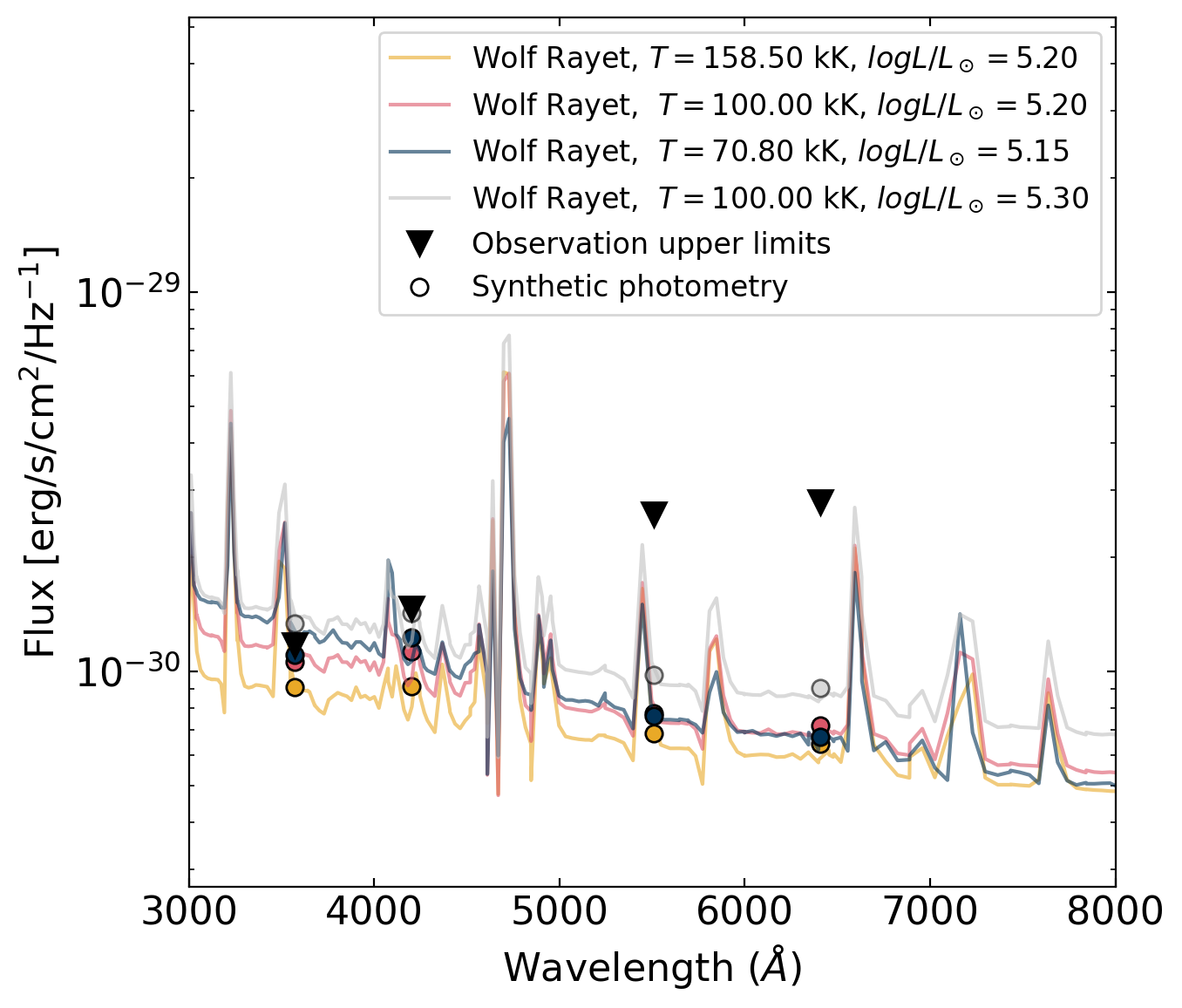}
    \caption{Example spectra of Potsdam WR models allowed by the upper limits of pre-explosion images taken by Large Binocular Telescope. Over a wide range of temperatures, we rule out WR models with log$(L/L_\odot)>5.2$ (e.g., the gray spectrum in this figure). }
    \label{fig:pre_explosion_SEDs}
\end{figure}

Using the pre-explosion photometry limits set by Large Binocular Telescope observations \citep{johnson2017}, we place constraints on the progenitor. 
We take a set of early nitrogen-sequence WR models with a range of temperatures from the Potsdam group \citep{Todt2015, Hamann2004}. Then, we scale the spectral models to different luminosities and perform synthetic photometry, which we compare to the observed limits. Between temperatures 40 kK and 170 kK, we can consistently rule out WR stars with log$(L/L_\odot)>5.2$ because of the $U$-band upper limit. Figure \ref{fig:pre_explosion_SEDs} shows example model spectra allowed (in yellow, pink, dark blue) and ruled out (in gray) by the pre-explosion upper limits. 

Using the relationship provided by \citet{Grafener+2011}, we can translate this luminosity limit into a limit on the pre-explosion mass of the progenitor star. Using their Eq. 10 for pure helium stars we find that log$(L/L_\odot)<5.2$ corresponds to $M_{\rm prog}< 10.56$\,\Msun. This limit is relatively high compared to inferences for most Type Ib/c SNe based on their ejecta masses \citep{Drout+2011,Taddia+2018,Barbarino+2021,Karamehmetoglu+2023}. If the progenitor were on the high end of the allowed range, this could indicate wind-driven stripping could have played a role in the removal of the remaining helium-rich layers prior to core collapse, but these limits make the likelihood of a single-star progenitor low.

\subsection{Constraints on Progenitor Age from Recent SFH}\label{sec:age}

\begin{figure*}
  \centering
  \includegraphics[width=6.5in]{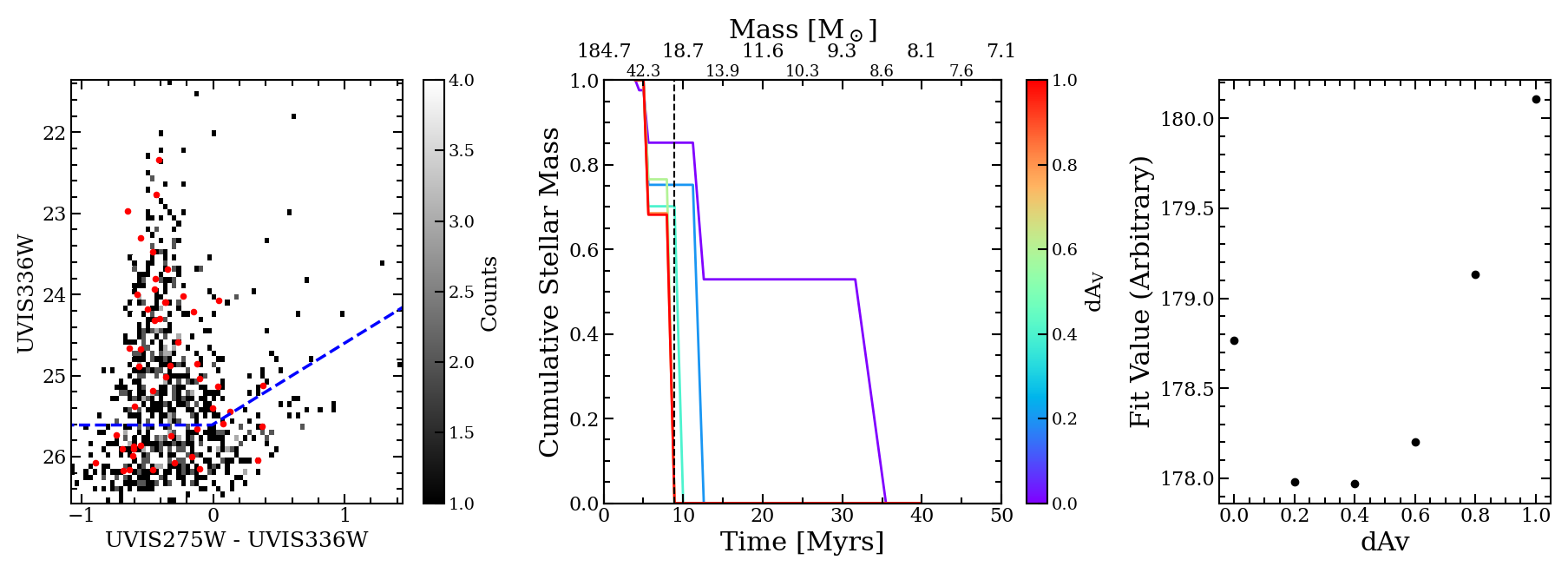}  
  \caption{Data and analysis used to constrain the differential reddening for the vicinity of SN~2012fh. \textit{Left:} Observed CMD within our region of interest. Red points indicate the stars within 50\,pc of the SN while the background field populations are plotted in gray scale and the magnitude limits of the data are shown as blue dashed lines.  \textit{Center:} The best-fit SFH for each tested differential extinction model. The cumulative fraction of stellar mass in each age bin is indicated by each line, and the different colors represent different amounts of modeled differential extinction.  Both of the best-fitting models ($dA_V=0.2$ and $dA_V=0.4$\,mag) have very similar results.  Higher amounts of differential extinction yield slightly younger best fits. \textit{Right:} The fit value for the best-fitting model for each $dA_V$ test, showing that the best-fitting model (by Poisson maximum likelihood) was $dA_V=0.4$\,mag, and adding more differential extinction to the model only worsens the fit.  
  }
\label{fig:dav}
\end{figure*}

The results of our SFH-fitting optimization are shown in Figure \ref{fig:dav}, including the impact of optimizing the extinction and differential extinction as well as the CMD from within 50\,pc of the SN location and the surrounding field.  We show only the past 50\,Myr under the assumption that the progenitor of the core collapse was a massive star with a short lifetime. The left panel shows the cumulative fraction of the population as a function of stellar age for the best-fit model from six different assumed values for differential extinction ($dA_V$, see Section~\ref{sec:fitting}).  The middle panel shows the fit value for these fits, where lower values indicate that the model is a better representation of the data. Our best fit occurs with $dA_V=0.4$\,mag.  The $A_V$ value of that fit is 0.05\,mag.  The fit with $dA_V=0.2$\,mag is of very similar quality, but also has a nearly identical age distribution, so our specific choice of $dA_V$ in this case does not impact the resulting age constraints.  However, we note that without the additional differential extinction included in the fitting ($dA_V=0$), the model includes a 30 Myr old population to fit the width of the MS. These extinction properties suggest that the region has only a few tenths of a magnitude of extinction beyond the estimate of 0.09\,mag from the all-sky extinction maps \citep{schlafly2011}.   This differential extinction is due to the stars residing at different depths in the dust near the SN location.

\begin{figure*}
  \centering
  \includegraphics[width=6.5in]{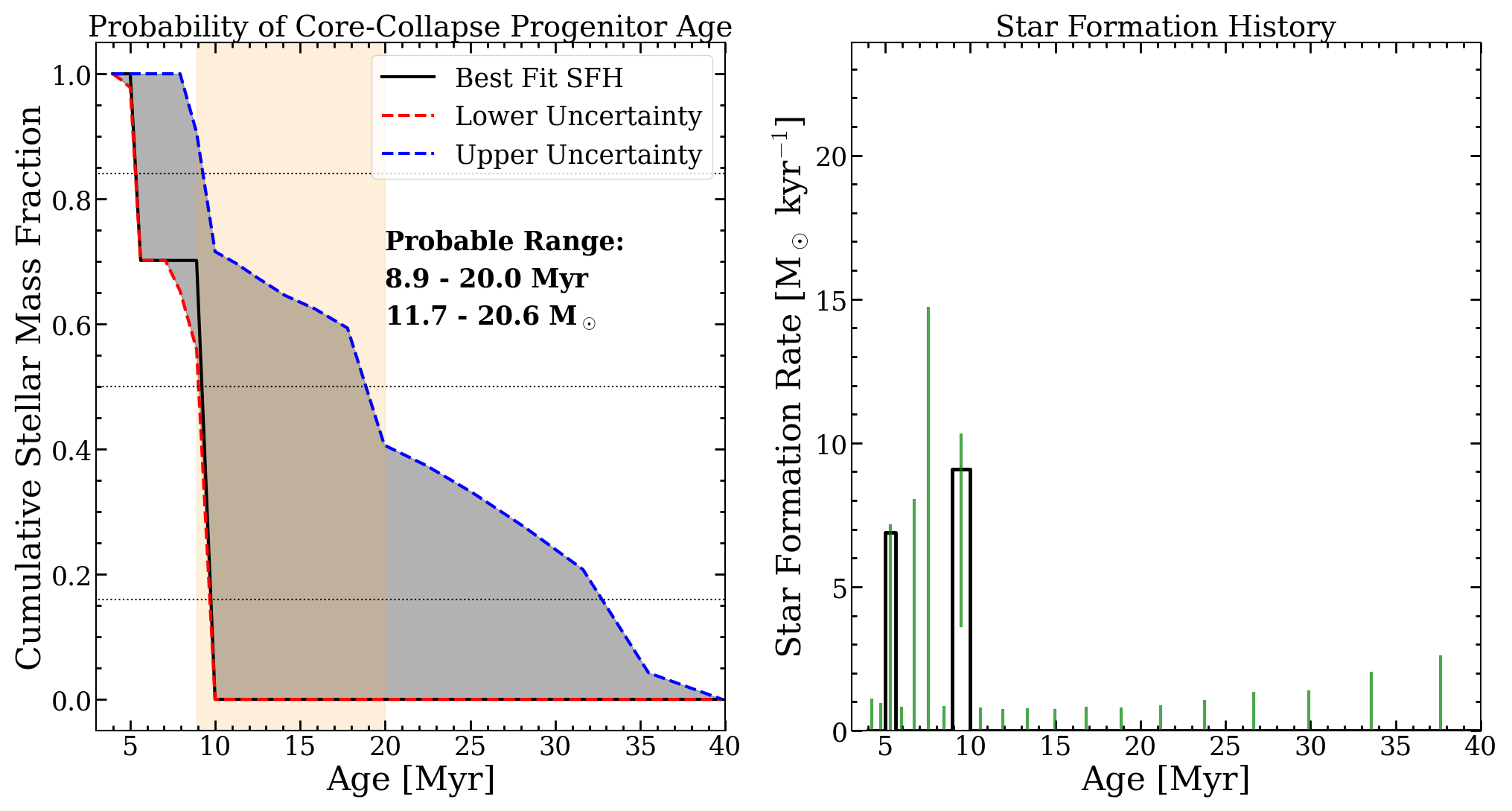}  
  \caption{Data and analysis used to constrain the progenitor age for SN~2012fh. \textit{Left:} The best-fit SFH and associated uncertainties produced by \texttt{hybridMC}. The cumulative fraction of stellar mass in each age bin is indicated by the red line. The gray shaded region depicts the 1$\sigma$ uncertainties in the SFH. The tan region shows the probable age range, which is taken to be the median population of the 1$\sigma$ uncertainties. \textit{Right:} The differential SFH, showing the star-formation rates as a black histogram and uncertainties (green error bars) that result in the gray shaded region of the cumulative fraction plot at left. 
  }
\label{fig:cdf}
\end{figure*}

Our best-fit recent SFH of the SN~2012fh location with optimized extinction parameters and full uncertainty estimations is provided in Figure~\ref{fig:cdf}.  The left panel displays the cumulative age distribution, which helps to show the impact of the covariance between neighboring age bins.  The right panel illustrates the best-differential SFH with the errors on each bin. These uncertainties clearly show that the presence of stars with ages from 5--10\,Myr is generally consistent with the data.  This SFH allows us to calculate the fraction of stars in the region at each of the fitted ages along with their uncertainties, and we find that the median age of the stars is 10$^{+10}_{-1}$\,Myr.  Thus, the progenitor was most likely formed in that time period. This median age does not fully capture the SFH, however, because there is some additional star formation at 5\,Myr, with high upper limits in between, which suggests the presence of some stars younger than 10\,Myr.  

These age constraints suggest the most likely (most prevalent population) single-star progenitor had an initial mass between 9\,\Msun\ and 20\,\Msun, but a more massive progenitor is possible because of the population with an age of $\sim$5~Myr, which may contain 30\% of the stellar mass with ages $<$40 Myr. \citet{maund2018} studied the population of stars surrounding SN~2012fh based also on PARSEC models and found ages from 4 to 16\,Myr, consistent with our SFH. \citet{sun2023} also measured ages for stars in this region, finding populations at 3 and 30\,Myr.  Our best fit has no star formation for all ages between 10 and 40\,Myr (nothing significant is detected, resulting in upper-limits on the star formation rate during those epochs), making our result more consistent with that of \cite{maund2018}. 

The lack of hydrogen and helium in the event spectrum makes a single-star progenitor in the 9--20\,\Msun system highly unlikely, since for stars with initial masses of around 20\,\Msun\ or less, stellar winds of single stars are not strong enough to remove the H envelope or the He layers \citep{smith14}.  Thus in order for the progenitor to be from a single star, it would most likely belong to the smaller, $\sim$5\,Myr old population.  However, the pre-supernova luminosity limits discussed in the previous subsection make that single-star scenario also unlikely.

Note that these age results assume single-star evolution models, although they do take into account binary companions that do not affect stellar evolution (such as long-period binaries) which evolve as two single stars but appear as a single point source in our data. 
While massive interacting binary stars would not evolve along such models, their age distribution should be similar to the other stars in the region since star formation typically occurs in clusters \citep[e.g.,][]{lada2003}.

\subsection{Constraints on the Progenitor Core Mass from Post-Explosion Spectroscopy}\label{sec:spectral_fits}
The amount of oxygen synthesized over the course of a star's lifetime is proportional to the helium core mass and can therefore be directly related to the preSN mass \citep{woosley1995}.
More massive stars synthesize more oxygen and are expected to have higher [\ion{O}{1}] $\lambda\lambda6300$, 6364 luminosity \citep{2011Elmhamdi, 2012Maguire}.
We use the non-local thermodynamic equilibrium radiative transfer code \texttt{CMFGEN} \citep{1998Hillier, 2013Dessart, 2019Hillier} models of \citet{2023Dessart} to model the relationship between [\ion{O}{1}] $\lambda\lambda6300$, 6364 luminosity and preSN mass to derive the preSN mass of SN~2012fh.
As input to \texttt{CMFGEN}, we use models of evolved He stars from \citet{2019Woosley} and \citet{2020Ertl} with initial He core masses between 2.6 and 12\,\Msun\ (see Table~\ref{tab:spec}), and use the mass-loss prescription of \citet{yoon2017}. 
To mimic stripping due to binary evolution, the hydrogen envelope of these stars is artificially removed prior to evolution.  This causes the mass loss from stellar winds to produce a preSN star with a significantly lower core mass than would be expected from single-star evolution \citep{2021Dessart}. 

We compare our spectra taken on days 135, 147, and 155 to models at days 133, 146, and 161, respectively. The models are empirically scaled to the observations using the ratio of the integrated flux of the observations to the models over the full range of the observed spectra with reasonable signal. In practice we integrate over 4000--7000\,\AA\ for the CCDS spectrum, 4000--9000\,\AA\ for the Kast day 147 spectrum, and 4500--9000\,\AA\ for the Kast day 155 spectrum, although we find our results are not particularly sensitive to the wavelength range used.

We evaluate each model using a reduced $\chi^2$ around the [\ion{O}{1}] doublet in the range 6050--6800\,\AA.
In all spectra, the minimum reduced $\chi^2$ occurs at $M_{\rm preSN}=5.63$\,\Msun; however, the minimum reduced $\chi^2$ is not 1, indicating that our uncertainties (at least around the [\ion{O}{1}] line) are underestimated.
We scale the flux uncertainties of the Kast spectra by a factor of 14 for day 147 and 3.6 for day 155 to obtain a minimum reduced $\chi^2$ of 1 for the $M_{\rm preSN}=5.63$\,\Msun\ model. 
For the day 135 CCDS spectrum, we find the best uncertainty is 18\% of the flux.
Table~\ref{tab:spec} lists the model initial masses, preSN mass, and reduced $\chi^2$ value at each epoch.
With these values we find that, in addition to the $M_{\rm preSN}=5.63$\,\Msun\ model, the $M_{\rm preSN}=7.24$\,\Msun\ model produces a reduced $\chi^2$ of 1.3--1.7 (depending on the epoch), while all other models ($M_{\rm preSN}<5.63$\,\Msun) yield reduced $\chi^2$ values $\geq4.4$.

\begin{figure*}
  \centering
  \includegraphics[width=\textwidth]{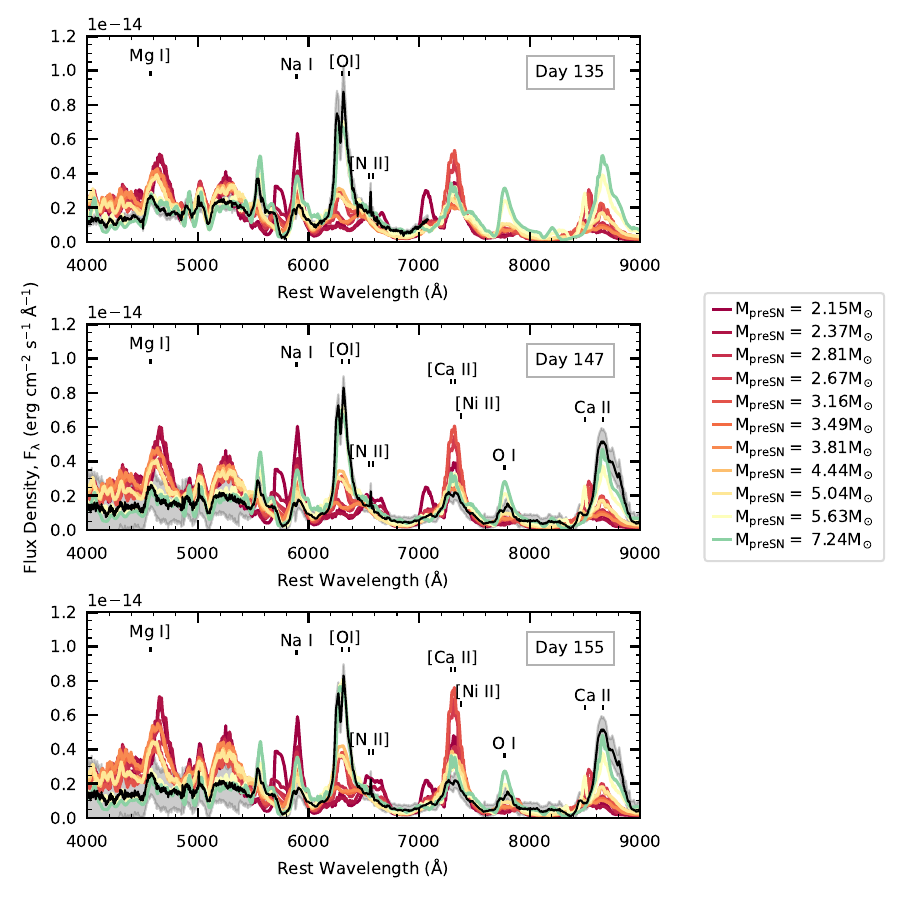}  
  \caption{Fits of SN spectral models, with preSN masses ranging from 2.1 to 7.2\,\Msun\ (colors), to the spectra of SN~2012fh taken 135 (top), 147 (middle), and 155 (bottom) days after the explosion (black). The inflated error bars on the observed spectrum are shown as the gray shaded region (the original errors are smaller than the width of the black line). Prominent features are labeled at their rest wavelengths.
  }
\label{fig:spectral_fits}
\end{figure*}

The results from our spectral fitting are provided in Table~\ref{tab:spec}, and example fits to the observed spectra are shown in Figure~\ref{fig:spectral_fits}.   While several models of late-time Type Ic SNe are consistent with the spectra, the consistent models all have relatively high core masses ($>5.6$\,\Msun), mostly because of the strong [\ion{O}{1}] $\lambda\lambda6300$, 6364 emission lines, which require masses this large.  Additionally, there are several other features that appear to fit the data more accurately for higher masses such as the red Ca II features. These high-mass spectral models are most consistent with the data, and even the most massive models available do not produce features that are inconsistent with our spectra.  Therefore, we can only provide a lower limit to the final core mass prior to the supernova.
Such a massive core would most likely come from a very large initial mass.  For example, the progenitor models of \citet{woosley1995} suggest such a $\sim$7\,\Msun preSN mass ($\sim$11\,\Msun He core mass at He ignition) would require a 30--35\,\Msun initial MS mass at solar metallicity. 

\begin{deluxetable*}{cccccccc}
\tablecaption{Post-explosion spectral modeling results,\label{tab:spec} showing the reduced $\chi^2$ when errors are inflated to make the minimum $\chi^2=1$ and with the errors as calculated from the data reduction.}
\tablehead{\colhead{Initial He }                   & \colhead{preSN}        & \colhead{$\chi^2$ day 135}            & \colhead{$\chi^2$ day 147}           & \colhead{$\chi^2$ day 155}           & \colhead{$\chi^2$ day 135} & \colhead{$\chi^2$ day 147} & \colhead{$\chi^2$ day 155}\\
           \colhead{Mass\tablenotemark{a} (\Msun)} & \colhead{Mass (\Msun)} & \colhead{($\chi^2_{\mathrm{min}}=1$)} & \colhead{($\chi^2_{\mathrm{min}}=1)$} & \colhead{($\chi^2_{\mathrm{min}}=1$)}& \colhead{}   & \colhead{}  & \colhead{}}
\startdata
2.6 & 2.15 & 11.3 & 19.4 & 23.5 & 0.4 & 3794 & 307\\
2.9 & 2.37 & 10 & 18.3 & 22.7 & 0.36 & 35938 & 296\\
3.3 & 2.67 & 5.6 & 9.4 & 10.8 & 0.26 & 2749 & 231\\
3.5 & 2.81 & 7.4 & 14.0 & 17.7 & 0.20 & 1845 & 141\\
4.0 & 3.16 & 4.5 & 7.8 & 9.1 & 0.16 & 1536 & 118\\
4.5 & 3.49 & 7.7 & 14.4 & 18.7 & 0.27 & 2829 & 243\\
5.0 & 3.81 & 7.9 & 14.2 & 18.1 & 0.28 & 2777 & 236\\
6.0 & 4.44 & 4.4 & 6.0 & 6.8 & 0.16 & 1182 & 88\\
7.0 & 5.04 & 4.8 & 7.5 & 8.7 & 0.17 & 1478 & 114\\
8.0 & 5.63 & 1.0 & 1.0 & 1.0 & 0.04 & 195 & 13\\
12.0 & 7.24 & 1.7 & 1.3 & 1.3 & 0.06 & 247 & 17\\
\enddata
\tablenotetext{a}{He-core mass at the time of central He ignition.}  
\end{deluxetable*}

\subsection{Constraints on a Surviving Companion from Post-Explosion Photometry}

In Figure~\ref{fig:post_color}, the lower-right panel shows the combined image from our 2022 WFC3/UVIS F275W and F336W data covering the location of SN~2012fh.  We determined the location of SN~2012fh, marked with the arrow in the top-left panel of Figure~\ref{fig:post_color}, using archival {\it HST} data.  As shown in the left panel, we detected the fading SN in data taken in 2014 (GO-13364).  The SN was no longer detected in data taken in 2016 (GO-14762) in F300X and F475X. Our photometry of these data showed no statistically significant detection of a remaining point source at the location of SN~2012fh.  Our AST results (average 5$\sigma$ detections) from the crowded-field photometry supplied initial upper limits of F275W $>$ 25.6 and F336W $>$ 25.9\,mag.   These upper limits are quite conservative, as they are consistent with those from the 2016 data and with the measured noise level at the location in our UV data, which was F275W = 28 and F336W = 29\,mag. Dedicated upper limits at the location using {\tt space\_phot}, derived directly from the variance in the pixels within the PSF core, yields F275W $>$ 26.38 and F336W $>$ 26.55\,mag, which we adopt as our final limits, being between our typical 5$\sigma$ detections and the noise level. 

\begin{figure*}
  \centering
  \includegraphics[width=\textwidth]{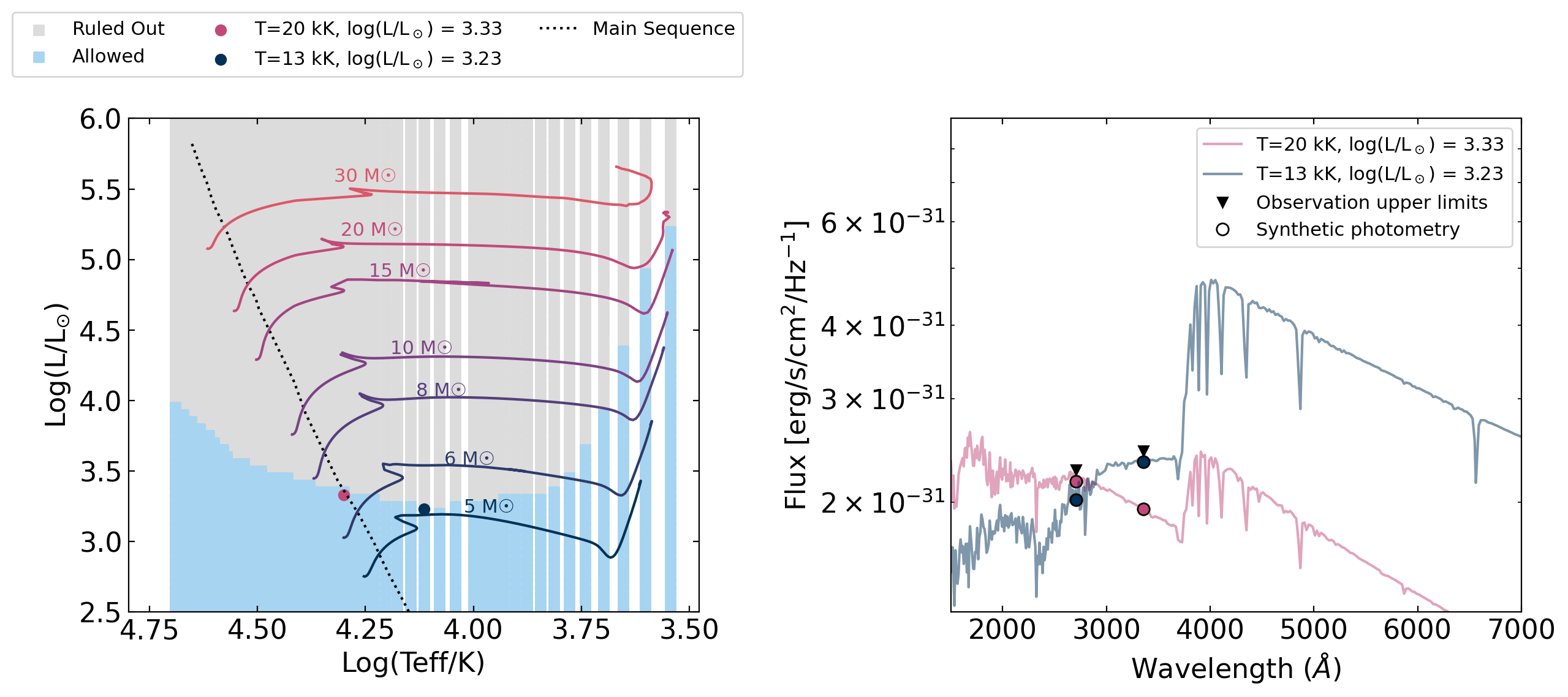}  
  \caption{\textit{Left:} The positions of potential stellar companions in the Hertzsprung-Russell (HR) diagram, allowed and ruled out by our photometric upper limits, assuming a \cite{1993Kurucz} H-rich model. We assume solar metallicity, and $\log\,g$ is set to 4 at temperatures below 36,000\,K and to 5 at temperatures above 36,000\,K. The single-star evolution predicted by \texttt{POSYDON} is shown for different masses through temperature-luminosity space, and the colors in the background indicate areas of temperature and luminosity space allowed by our nondetections. For MS companions (dotted line), we rule out log($L/L_\odot)>3.35$. Because our deepest data are in blue and UV bands, our mass limits are most constraining for less-evolved hot stars. \textit{Right:}  Spectral fits to the upper limits of our {\it HST} photometry. The pink spectrum represents MS stars with the most constrained luminosity, corresponding to the pink point of the HR diagram in the left panel. The blue spectrum represents stars with the tightest constraint on luminosity at 13,000\,K, which is located at the dark blue point in the left panel. The triangles indicate the upper limits of our observations, which are compared to the synthetic-photometry outputs from the model spectra. }
\label{fig:SED}
\end{figure*}

With our post-explosion upper limits, we put constraints on any surviving companion on the temperature-luminosity space, shown in the left panel of Figure \ref{fig:SED}. As we expect any surviving companion to be hydrogen-rich, we implement the \cite{1993Kurucz} H-rich stellar atmosphere models. At each available temperature in the Kurucz models, we scale the spectrum so that it integrates to reach a targeted luminosity, ranging between $\log\,(L/L_\odot)$ of 2.5 and 6 (i.e., the range of the luminosity axis). We then simulate an observation of this star by correcting for both the distance and line-of-sight extinction to SN~2012fh (see Sections ~\ref{sec:intro} and \ref{sec:age}) using the \citet{Cardelli1989extinction} extinction curve. Then, we apply synthetic photometry on the final spectrum to obtain magnitudes, which we compare to the observed upper limits. If the calculated magnitudes are lower than our upper limits (i.e., brighter), we rule out this temperature and luminosity combination. Throughout our analysis, we assume solar metallicity. We use models with $\log\,g = 4$ for most of the grid, with the exception of higher temperature models ($T > 36,000$\,K) where we use $\log\,g = 5$. 

Between temperatures of 6000 and 26,000\,K, we can constrain the luminosity to be $\log\,(L/L_\odot) < 3.45$. Any surviving MS companion (along the dotted line in Figure~\ref{fig:SED}) would have $\log\,(L/L_\odot) < 3.35$ at 20,000\,K. We find that the luminosity has the deepest constraint of $\log\,(L/L_\odot) < 3.25$ at $\sim 13,000$\,K. In the right panel of Figure \ref{fig:SED}, we show two examples of stellar spectra in comparison with our observed upper limits. The pink spectrum represents MS stars at the boundary of allowed and ruled-out regimes. The blue spectrum represents stars with the tightest luminosity constraint at 13,000\,K.

Solar metallicity single-star evolution tracks from \texttt{POSYDON} (described in Sec.~\ref{sec:discussion}) of different stellar masses are overplotted in the left panel of Figure \ref{fig:SED}, showing that massive, evolved stars are also possibilities at the cooler temperatures. Because our deepest data are in blue and UV bands, our mass limits are most constraining for less-evolved hot stars. Any MS companion is likely to be $< 8\,M_\odot$, the lowest such mass limit for an SN~Ic progenitor companion to date. 
Since the companion is typically the initially less massive secondary, it is mostly expected to be  on its MS owing to its longer evolutionary timescale compared to the progenitor. Synchronized post-MS evolution is rare, largely because it occurs only in near-equal-mass binaries, making evolved giant companions to SESNe an uncommon prediction \citep[$\lesssim$5\% of SESNe;][] {zapartas2017b,Zapartas+2025}. Our luminosity upper limits will be examined in more detail in the context of binary population synthesis models in Section~\ref{sec:discussion}.

\section{Discussion} \label{sec:discussion}

\begin{figure*}
  \centering
  \includegraphics[width=0.49\linewidth]
{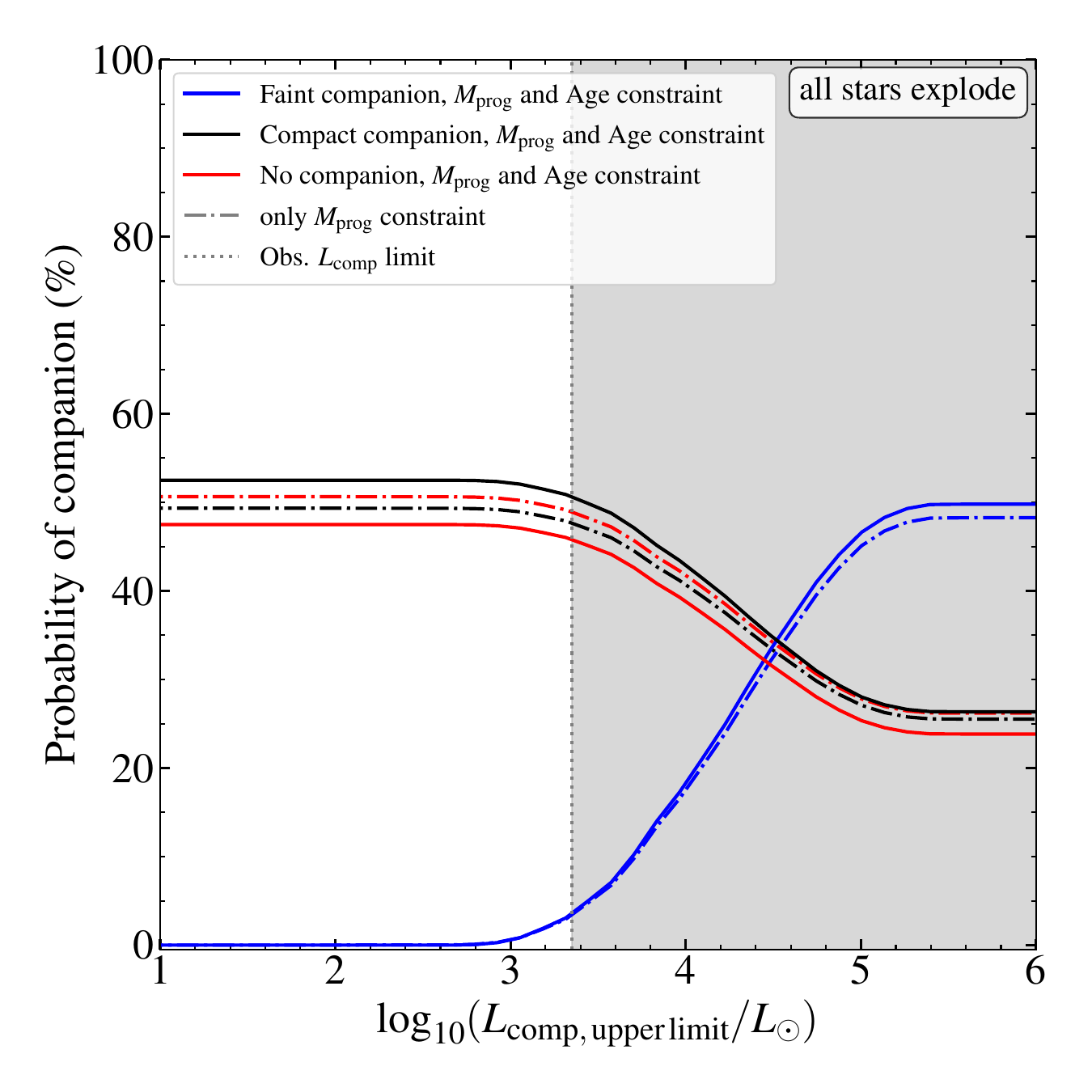}
  \includegraphics[width=0.49\linewidth]
{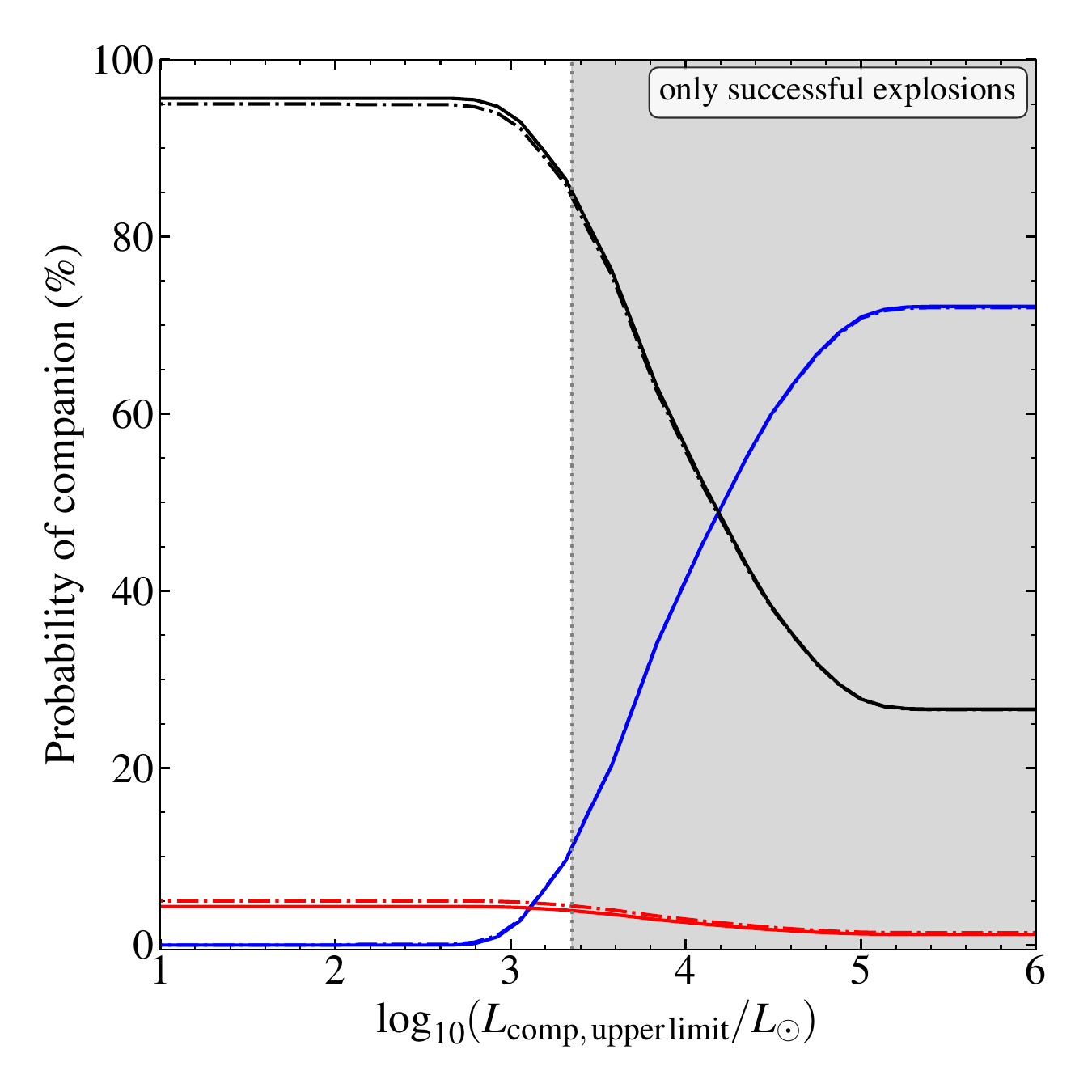}
  \caption{The probability of a stellar companion below our luminosity limits (blue), nondetectable compact-object companion (black), and companion absence (red) at the site of SN~2012fh according to our \texttt{POSYDON} population-synthesis models, given the luminosity constraints on a possible luminous companion. The gray shaded portion is ruled out by the upper limits on a possible companion luminosity in this work. At left, we assume all massive core-collapse progenitors to result in a successful explosion, whereas (right) we accept only the successful explosions. We show the case of no age constraint for the SN (dotted-dashed line), and also including an age constraint of 8--20\,Myr from the surrounding region (solid line).  
  }
\label{fig:prob_faint_comp_withAgetoo}
\end{figure*}

In this work, we combine four independent observational constraints for the progenitor system of SN~2012fh: (i) the age of the progenitor system, inferred from its surrounding population ($<$20\,Myr),  (ii) the progenitor core mass limits from pre-explosion photometry ($M_{\rm He}<10.56$\,\Msun), (iii) the progenitor core mass ($M_{\rm He}>5.6$\,\Msun), and (iv) deep limits on the luminosity of a possible MS binary companion at the moment of explosion, $\log\,(L_{\rm comp}/L_{\odot}) \lesssim 3.35$. The simultaneous existence of these constraints on the SN progenitor makes SN~2012fh a special event.

To elucidate the nature of the progenitor system,  we employ a binary population synthesis framework to compare our model predictions of progenitor age, mass, and, where applicable, companion luminosity at the time of the explosion for SNe Ib/c with the observational constraints derived above.
For this, we used  v2 of the binary population synthesis code \texttt{POSYDON}\footnote{\href{https://posydon.org/}{posydon.org}. Exact commit used in this work is \href{https://github.com/POSYDON-code/POSYDON/tree/Dimitris_WD_mergers}{891c5897}.}   \citep{fragos2023, Andrews+2024} which is based on extensive grids of detailed single and binary evolution  models computed with \texttt{MESA} \citep{paxton2011,paxton2013, paxton2015, paxton2018, paxton2019}\footnote{\texttt{POSYDON}~v2 grids will be archived in the \href{https://zenodo.org/communities/posydon/}{\tt POSYDON} Zenodo community.}, and accounting for all major possible evolutionary scenarios (albeit with a simplified treatment for merger evolution) toward SN~Ib/c  progenitors \citep[defined as massive stars that eject $<0.033$\,\Msun\ hydrogen mass in the explosion;][]{gilkis2019}. These evolutionary pathways encompass progenitors stripping through stable mass transfer, common-envelope evolution, or stripping through solely stellar winds in the case of isolated progenitors, as well as scenarios involving mergers or disruption following the final fate of one component in a binary system. For a detailed description on the code and default assumptions see \citet{Andrews+2024}. Here, we summarize the key assumptions in our work. \texttt{POSYDON}, built on detailed stellar models, allows improved treatment of mass-transfer stability, and adopts a rotationally limited accretion. The models follow default wind prescriptions \citep{Andrews+2024}, largely based on the \texttt{MESA} ``Dutch’’ scheme. Unstable mass transfer leads to common-envelope evolution, modeled using the $\alpha_{\rm CE}$–$\lambda_{\rm CE}$ formalism \citep{Webbink1984, Livio+1988}, with $\alpha_{\rm CE} = 1$, $\lambda_{\rm CE}$ computed from the donor’s stellar structure, and with the core-envelope boundary set where the hydrogen fraction drops below 0.3. For binary merger scenarios, we slightly depart from the default: if one component has left the main sequence, we match to a single-star model as an approximation for the merger product,  matching only its carbon-oxygen or helium core and its mass-weighted helium abundance, an approach that improves its ability to produce an explosion while leaving the envelope mass of the merger product, and hence the resulting SN type, more uncertain.

We evolved 2 $\times 10^{5}$ stellar systems at solar metallicity \citep[using initial-final interpolation;][]{fragos2023}, considering initial masses of $M_1$ between 4 and 250\,\Msun for both the initially more massive stars in binary systems and single stars, adopting a binary fraction of 0.6 within
the orbital period range relevant for interactions ($\log_{10}(P/\mathrm{d}) \le 3.5$) \citep[e.g.,][]{sana2012,Moe+2017,Banyard+2022}. 
With this setup, we computed the probability and bolometric luminosity $L_{\rm comp}$ of companions, where present, next to SN~Ib/c progenitors. An extensive study of the properties of companions next to SESNe is presented by \citet{Zapartas+2025}, but here we focus on the case of SN~2012fh. 
To evaluate the significance of our constraint on a potential companion, Figure~\ref{fig:prob_faint_comp_withAgetoo} presents the probability associated with each evolutionary scenario for SN~2012fh.


We constrain the progenitor mass to the range $5.6 < M_{\rm prog}/$\Msun$< 10.56$, where the lower limit is set by our post-SN spectroscopic analysis (Section~\ref{sec:spectral_fits}) and the upper limit by the nondetection of a luminous WR star in pre-explosion imaging (Section~\ref{sec:pre_explosion}).  The companion's nondetection, going as deep as $\log\,(L_{\rm comp}/L_{\odot}) \lesssim 3.35$, excludes the presence of a MS star at the SN site of $\sim 7$--8\,\Msun or above (at $Z_\odot$). The combined companion constraint (vertical shaded region in Figure~\ref{fig:prob_faint_comp_withAgetoo}), and the relatively high progenitor mass of SN 2012fh together are sufficiently stringent to rule out the presence of almost any faint stellar companion. This makes SN 2012fh an intriguing case, as the commonly expected origin process of binary stripping onto the initially less massive stellar companion \citep[e.g.,][]{zapartas2017b} appears to be excluded. 

We note that low-luminosity companions (log$(L/L_\odot) \lesssim 2.5$) are generally uncommon in our models \citep{Zapartas+2025} because mass transfer between a donor star massive enough to collapse ($\gtrsim 7 \Msun$) and a low-mass, low-luminosity companion tends to be unstable.  Such unstable mass transfer leads to a common envelope and eventual merging (no SESN with a companion) unless an exceptionally efficient envelope ejection is assumed, corresponding to $\alpha_{\rm CE} > 1$ \citep{Ivanova+2013,Fragos+2019,Klencki+2021}.  Moreover, specifically for SN 2012fh, the initial mass of the progenitor was more than the threshold of 7 \Msun (as in the end it formed a stripped star in the range 5.6--10.56\,\Msun at explosion). In our binary synthesis modeling, we find that the more massive the progenitor, the more massive and luminous its companion is expected to be, for the same argument as above. Thus, we argue that the deep limits at the SN 2012fh site are sufficient to exclude almost all possible stellar companions. Even deeper observational limits would only modestly strengthen this conclusion.

We consider two cases: one in which all massive core-collapse progenitors explode (Figure~\ref{fig:prob_faint_comp_withAgetoo}, left panel), and another that includes only those progenitors that successfully explode according to the criteria of \citet{Ertl+2016} and \citet{Patton+2020}, thereby excluding most massive WR stars as possible SN~Ib/c progenitors \citep[right panel; see also][]{zapartas2021b}. The probability of a stellar companion below our luminosity limit is ${\sim}3\%$ and $9\%$, for everything exploding or accepting only successful SNe, respectively.  The limited amount of models allowed is because the progenitor is sufficiently massive for its stellar winds to play a significant role in its stripping, which makes the presence of a companion star not strictly required \citep[although observational evidence for reduced mass loss would weaken this argument; e.g.,][]{smith14}.

As faint stellar objects are unlikely, the progenitor scenario, under the assumption that all massive stars explode, bifurcates into either a $\sim 51\%$ probability of having no companion at the SN site, or a $\sim 46\%$ probability of a compact-object companion. In contrast, when only successful explosions according to \citet{Ertl+2016} and \citet{Patton+2020} are considered, the compact-object companion scenario becomes strongly favored, with a probability of $\sim 86\%$. In most cases of compact-object companion, it is expected to be a BH that remained bound to the binary system following its formation, owing to its low kick during collapse (following \citealt{Andrews+2024}). In this scenario, SN~2012fh would originate from the initially less massive secondary star, which was stripped of its envelope through mass transfer onto the compact object, potentially rendering the progenitor system an X-ray-bright binary prior to explosion. While our analysis groups Type Ib/c SNe together, it remains unclear how progenitors can evolve into an SN~Ic through purely binary mass transfer and subsequent wind stripping of the helium envelope, with common-envelope scenarios involving the BH to be likely. SN 2012fh cannot be classified as an ultrastripped SN \citep{Tauris+2015} owing to its relatively high ejecta mass. Including the age constraint for the progenitor of $<20$\,Myr does not significantly alter the probability of a compact-object companion at the SN site. However, it increases the likelihood that such a companion would be a BH, as more massive and shorter lived systems are more prone to BH formation.

The remaining probability for the absence of a companion is $<5\%$ in the case of only successful explosions (red line on the right panel of Figure~\ref{fig:prob_faint_comp_withAgetoo}). Although, in principle, such progenitors could originate from either true single stars or from stars ejected during binary disruption, our results instead favor a binary merger origin for them \citep{de-Mink+2014}. Nevertheless, the limitations of our merger modeling prevent us from drawing firm conclusions on this origin scenario. Progenitors formed through these channels are expected to explode on longer timescales than single WR stars \citep{Zapartas+2017}, having lost their outer envelopes primarily due to binary interaction, and subsequently shedding their remaining helium-rich layers via stellar winds during their final evolutionary phases as stripped stars. Weaker wind mass loss would further reduce the possibility for this scenario \citep[e.g.,][]{smith14, Beasor+2020, Antoniadis+2024, Antoniadis+2025}.

Note that the exclusion of a faint stellar companion in the case of SN 2012fh, unlike the general expectation for companions in SESNe \citep{Zapartas+2025}, is not particularly sensitive to assumptions about how conservative stable binary mass transfer is. 
More conservative mass transfer would predict more massive and thus more luminous stellar companions, which are not observed in the case of SN 2012fh, thereby increasing the tension with the nondetection. Consequently, the probabilities derived from our default \texttt{POSYDON} analysis, characterized by fairly nonconservative mass transfer regulated by rotationally limited accretion (although for short orbits and high masses, the average efficiency may be higher; \citealt{Andrews+2024}), should be regarded as optimistic estimates for the presence of an undetected stellar companion. Under more conservative assumptions, the absence of a luminous companion would become even more significant, further supporting the conclusion that no stellar companion was present at the time of explosion. More generally, a population of companion constraints can provide valuable insight into the physics of binary interaction, disfavoring fully conservative mass-transfer scenarios \citep{Zapartas+2025}, and serving as an independent test of the degree of mass-transfer conservation in both massive stellar and compact-object binaries \citep[e.g.,][]{de-Mink+2007,Vinciguerra+2020,Lechien+2025}.

\section{Conclusions}

We have synthesized previous studies of SN~2012fh with our own spectral and photometric analysis of the precise location and vicinity to place improved constraints on the physical characteristics of the progenitor system.  Our measurements suggest that the age of the system was most likely to be $<20$\,Myr.  The spectroscopy shows that SN~2012fh was an SN~Ic with a preSN He core mass of $>5.6$\,\Msun.  The photometry constraints on the location prior to the SN explosion also suggest a binary progenitor.  The photometry after the event limits the luminosity of any binary companion to $\log(L/L_{\odot}) \lesssim 3.35$.  

Using the binary evolution models of \texttt{POSYDON}, we are able to use the observed limits to constrain the model systems that could have produced such an event. A stellar companion (which is commonly expected in SESNe) is ruled out to $\lesssim 10\%$ probability. There is a high probability ($\gtrsim50 \%$) that there is a BH companion at the SN site, which becomes the dominant scenario ($86\%$) if we take into account the possibility that massive WR stars do not explode. In this scenario, SN~2012fh was the explosion of the initially less massive star in a binary, stripped after interaction with the BH companion. The remaining minority channels involve an isolated progenitor with no companion next to it at the time of explosion.

\bigskip

Support for this work was provided in part by grants GO-16165 and GO-17203 from the Space Telescope Science Institute (STScI) which is operated by AURA, Inc., under NASA contract NAS 5-26555. Additional support for A.V.F. was provided by the Christopher R. Redlich Fund and many individual donors.
E.Z. and D.S. acknowledge support from the Hellenic Foundation for Research and Innovation (H.F.R.I.) under the ``3rd Call for H.F.R.I. Research Projects to support Post-Doctoral Researchers'' (Project 7933). 
K.A.B. is supported through the LSST-DA Catalyst Fellowship project; this publication was thus made possible through the support of grant 62192 from the John Templeton Foundation to LSST-DA.
M.R.D. acknowledges support from the NSERC through grant RGPIN-2019-06186, the Ontario Ministry of Colleges and Universities through grant ER22-17-164, the Canada Research Chairs Program, and the Dunlap Institute at the University of Toronto.
D.M.\ acknowledges support from the U.S. National Science Foundation (NSF) through grants PHY-2209451 and AST-2206532. 
J.J.A. acknowledges support for Program number (JWST-AR-04369.001-A) provided through a grant from the STScI under NASA contract NAS5-03127. 
M.M.B. was supported by the Boninchi Foundation, the Swiss National Science Foundation (project \#CRSII5\_21349), and the Swiss Government Excellence Scholarship. M.K. was supported by the Swiss National Science Foundation Professorship grant (PI Fragos, project \#PP00P2 176868). S.G., C.L., P.M.S., and E.T. were supported by the Gordon and Betty Moore Foundation (PI Kalogera, projects GBMF8477 and GBMF12341). 
The authors wish to acknowledge the helpful resource of the Supernova Database in observation planning (http://heracles.astro.berkeley.edu/sndb/info).

\software{CMFGEN \citep{1998Hillier, 2013Dessart, 2019Hillier}, lightcurve\_fitting \citep{2024Hosseinzadeh}, DOLPHOT\citep{dolphin2016}, MATCH\citep{dolphin2012,dolphin2013}
}
\bibliography{references}{}
\bibliographystyle{aasjournal}

\end{document}